%% file: transp.tex
\documentclass{article}
\usepackage{amssymb}
\usepackage[dvips]{graphicx}

\if@twoside m
\else
\fi
\def\omit#1{}                               

\newcommand{\Real}{\mathop{\mathbb R}\nolimits}         
\newcommand{\Nat}{\mathop{\mathbb N}\nolimits}          
\newcommand{\Int}{\mathop{\mathbb Z}\nolimits}          
\newcommand{\ie}{{\em i.e.\/}}                          
\newcommand{\eg}{{\em e.g.\/}}                          
\newcommand{\cf}{{\em cf.\/}}                           
\newcommand{\etc}{{\em etc.\/}}                         
\newcommand{\qed}
           {\mbox{\quad\rule[-1.5pt]{.4em}{1.5ex}}}     
\newcommand{\si}{\mathop{\boldmath L^{1}}\nolimits}     
\newcommand{\sii}{\mathop{\boldmath L^{2}}\nolimits}    
\newcommand{\Comp}                                      
           {\mathop{\boldmath C_{0}^{\infty}}\nolimits}
\newcommand{\sobi}                                      
           {\mathop{\boldmath W_{2}^{1}}\nolimits}
\newcommand{\sobii}                                     
           {\mathop{\boldmath W_{2}^{2}}\nolimits}
\newcommand{\im}{\mathop{\rm Im}\nolimits}              

\newcommand{\PF}{{\sc Proof:\quad}}                     
\newtheorem{claim}{Claim}[section]
\newtheorem{thm}[claim]{Theorem}                        
\newtheorem{lemma}[claim]{Lemma}                        
\newtheorem{rem}[claim]{Remark}                         

\begin{document}

\title{Quantum waveguides with a lateral semitransparent barrier:
       spectral and scattering properties}
\author{P. Exner$^{a,b}$ and
D. Krej\v{c}i\v{r}\'{\i}k$^{a,c}$}
\date{}
\maketitle
\begin{quote}
{\small \em a) Nuclear Physics Institute, Academy of Sciences,
25068 \v Re\v z near Prague \\
 b) Doppler Institute, Czech Technical University,
B\v rehov{\'a} 7, 11519 Prague, \\
 c) Faculty of Mathematics and Physics, Charles University,
V Hole\v{s}ovi\v{c}k\'ach 2, \\
\phantom{d) } 18000 Prague, Czech
Republic \\
 \phantom{e) } \rm exner@ujf.cas.cz, krejcirik@ujf.cas.cz}
 \end{quote}
\vspace{8mm}

\begin{abstract}
\noindent We consider a quantum particle in a waveguide which
consists of an infinite straight Dirichlet strip divided by a thin
semitransparent barrier on a line parallel to the walls which is
modeled by a $\delta$ potential. We show that if the coupling
strength of the latter is modified locally, \ie, it reaches the
same asymptotic value in both directions along the line, there is
always a bound state below the bottom of the essential spectrum
provided the effective coupling function is attractive in the
mean. The eigenvalues and eigenfunctions, as well as the
scattering matrix for energies above the threshold, are found
numerically by the mode-matching technique. In particular, we
discuss the rate at which the ground-state energy emerges from the
continuum and properties of the nodal lines. 
Finally, we investigate a system with a modified
geometry: an infinite cylindrical surface threaded by a
homogeneous magnetic field parallel to the cylinder axis. The
motion on the cylinder is again constrained by a semitransparent
barrier imposed on a ``seam'' parallel to the axis.
\end{abstract}

\section{Introduction}

\noindent Quantum mechanics of constrained systems is experiencing
a new wave of interest connected with the recent progress in
semiconductor physics: nowadays experimentalists are able to
investigate the behavior of electrons in structures of various
shapes, at times rather elaborated. The small size, extreme
material purity, and its crystallic structure make it possible to
derive basic properties of these systems in a crude but useful
model in which the electron is considered as a free particle (with
an effective mass) whose  motion is constrained to a prescribed
subset of $\Real^d$ with $d=2,3$, possibly in presence of external
fields.

On the theoretical side, this inspires questions about relations
between spectral and scattering properties of such systems and the
underlying geometry and topology. A class of systems which
attracted a particular attention are {\em quantum waveguides}, \ie\/
tubular regions supporting a Schr\"odinger particle. It is known
that a deviation from the straight tube can induce existence of
bound states and resonances in scattering, vortices in probability
current, \etc, be it bending \cite{DE1, DEM, DES, ES1, GJ}),
protrusion or a similar local deformation \cite{AS, BGRS, EV1},
waveguide coupling by crossing \cite{SRW}, or by one or several
lateral windows \cite{ESTV, EV2, EV3} (the related bibliography is
rather extensive; the quoted papers contain many more references).

\begin{figure}[!htb]
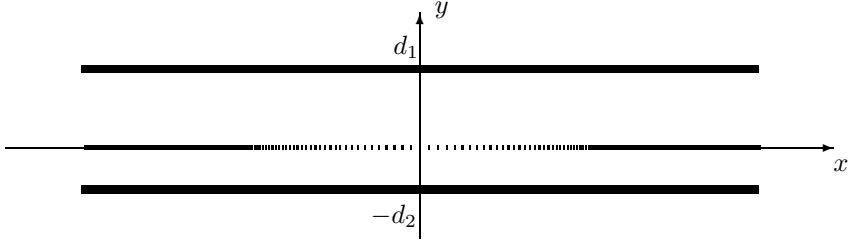

\begin{center}
\input fig01
\end{center}
\caption{Double waveguide with a $\delta$ barrier.}\label{schm}
\end{figure}

In this paper we are going to discuss a system closely related to
the last named one. It supposes again a double waveguide; however,
the coupling between the two parallel ducts will entail now a
tunneling through a thin semitransparent barrier rather than a
window in a hard wall separating them --- \cf~Figure~\ref{schm}.
To get a solvable model we describe the barrier by a $\delta$
potential whose coupling strength may vary longitudinally: the
Hamiltonian can be then formally written as
\begin{equation} \label{formal}
H_{\alpha}= -\Delta_{\Omega}+\alpha(x)\delta(y)
\end{equation}
with the barrier supported by the $x$--axis, where
$\Omega:=\Real\times(-d_2,d_1)$ is the double--guide strip.

There are several motivations to investigate a leaky--barrier
waveguide pair. First of all, it is a generalization in a sense of
earlier results, because the pierced--hard--wall case of
Ref.~\cite{ESTV} corresponds to $\alpha=0$ in the window and
$\alpha=\infty$ otherwise. Recall that the latter can serve to
describe an actual quantum--wire coupler --- see, \eg,~\cite{HTW,
Ku} --- and such a model will certainly become more realistic if
the tunneling through the barrier of a doped semiconductor
material separating the two guides is taken into account. At the
same time, the Hamiltonian~(\ref{formal}) covers for various
$\alpha$ a wide variety of situations.

On the mathematical side, the $\delta$ potential of~(\ref{formal})
can be treated more easily than the hard--wall barrier, since two
operators with different functions $\alpha$ have the same form
domain. To illustrate the difference, one can compare the
variational proof of existence of bound states in Thm~\ref{existence} below
with the analogous argument of Ref.~\cite{ESTV}. A deeper
application of the quadratic--form perturbations allows us to
construct the Birman--Schwinger theory for the waveguide systems
in question, in particular, to derive the weak--coupling behaviour
of the bound states. This will be done in a subsequent
paper~\cite{EK}.

Let us describe briefly the contents of the paper. In the next section
we shall describe the model and deduce its spectrum in the
``unperturbed'', \ie\/ translationally invariant case. 
In Section~\ref{Sec.Existence} we demonstrate that a local change of
the coupling parameter will cause the existence of bound states provided
it is negative in the mean. To ilustrate the spectral and also scattering
properties we shall discuss then in detail the example in which  
the barrier function is of a ``rectangular well'' shape. In the final
section we will show how the situation modifies if the semitransparent
barrier is placed at the surface of a cylinder threaded by a homogeneous
magnetic field.

\setcounter{equation}{0}
\section{Preliminaries}
\subsection{The Hamiltonian}
     Let $\Omega:=\Real\times{\cal O}$ with  ${\mathcal
O}:={\cal O}_{2}\cup{\cal O}_{1}:= (-d_{2},0)\cup(0,d_{1})$ be the
configuration space, \ie, the part of $\Real^2\,$ occupied by the
waveguide. Passing to the rational units, $\hbar=2m=1$, we may
identify the particle Hamiltonian $H_{\alpha}$ with the Laplace
operator away of the waveguide boundary and the barrier. To give
meaning to the formal expression (\ref{formal}) one has to specify
the boundary conditions. At the outer edges we assume the Dirichlet
condition,
\begin{equation} \label{Dirichlet}
 \psi(x,-d_{2})=\psi(x,d_{1})=0,
\end{equation}
while the barrier is transversally the usual $\delta$ potential
defined conventionally as
\begin{equation} \label{barrier}
 \psi(x,0+)=\psi(x,0-)=:\psi(x,0), \quad
 \psi_{y}(x,0+)-\psi_{y}(x,0-)=\alpha(x) \psi(x,0)
\end{equation}
for any $x\in\Real$ --- \cf~\cite[Sec~.I.3]{AGH} --- where the
subscript denotes partial derivative with respect to $y$. The
Hamiltonian domain is then
\begin{equation}\label{Hamiltonian}
  D(H_{\alpha}):=
  \left\{
  \psi\in\sobii(\Omega)| \;\: \psi\quad {\rm satisfies}\,
  (\ref{Dirichlet}) \;{\rm and}\; (\ref{barrier})\;
  \right\},
\end{equation}
where the function $\alpha:\Real\rightarrow\Real$, assumed to be
piecewise continuous, determines the shape of the barrier and
represents the $x$-dependent coupling ``constant'' of the
interaction.

For the sake of simplicity we shall exclude the above mentioned
case of a Dirichlet barrier, $\alpha(x)=\infty$ at a subset of
$\Real$. In that case all the operators $H_{\alpha}$ have the same
form domain, and the associated quadratic form is obtained by a
simple integration by parts:
\begin{eqnarray}
  q_{\alpha}[\psi] &:=&
     \int_{\Real\times{\bar{\cal O}}}|\nabla\psi|^2(x,y)dx dy+
     \int_{\Real}\alpha(x)|\psi(x,0)|^2 dx, \label{form}
     \\ \nonumber \\
  D(q_{\alpha})    &:=&
     \{\psi\in\sobi\left(\Real\times(-d_{2},d_{1})\right)|\
     \forall
     x\in\Real:\psi(x,-d_{2})=\psi(x,d_{1})=0\}.\label{formdomain}
\end{eqnarray}
The form (\ref{form}) is obviously symmetric and it is not
difficult to check that it is closed and thus indeed associated
with $H_{\alpha}$. Hereafter we adopt the notation of \cite{ESTV}:
\mbox{$d:=\max\{d_{1},d_{2}\}$}, \mbox{$D:=d_{1}+d_{2}$}, and
\[\nu:=\frac{\min\{d_{1},d_{2}\}}{\max\{d_{1},d_{2}\}}.\]
Without loss of generality we may assume that $d_{2}\leq d_{1}=d$.

\subsection{The unperturbed system}
        If $\alpha(x)=\alpha$ is a constant function, we can
solve the Schr\"odinger equation $H_{\alpha}\psi=k^2\psi$ by
separation of variables. To get the transverse eigenfunctions we
have to match smoothly the solutions in the two ducts,
\mbox{$C_{2}\sin \ell(y+d_{2})$} and \mbox{$C_{1}\sin
\ell(y-d_{1})$}, chosen to satisfy the condition
(\ref{Dirichlet}). If $\ell d_{1}, \ell d_{2}$ are not multiples
of $\pi$ we get thus the following condition on eigenvalues of the
transverse part of the Hamiltonian:
\begin{equation}\label{spc}
  -\alpha=\ell\,(\cot \ell d_{1}+\cot \ell d_{2}).
\end{equation}
\vspace{2mm}
\begin{rem}\label{rem1}{\rm
    If $d_{1},d_{2}$ are rationally related the Schr\"odinger
equation can be also solved by $\ell=\frac{\pi n
p}{d_{1}}=\frac{\pi n q}{d_{2}},\; n\in\Nat\setminus\{0\}$.
However, such wave functions are zero at $y=0$, and therefore
independent of $\alpha$. In this sense they represent a trivial
part of the problem. A prime example is the symmetric waveguide
pair, $d_{1}\!=\!d_{2}$, where this observation concerns every
solution antisymmetric w.r.t. $y\!=\!0$. It is reasonable to
concentrate on the nontrivial part only. If
$\nu\equiv\frac{d_{2}}{d_{1}}=\frac{p}{q}$, we denote by ${\cal
G}_{\nu}$ the subspace in $\sii(-d_{2},d_{1})$ spanned by the
solutions of (\ref{spc}). Putting then ${\cal
H}_{\nu}:=\sii(\Real)\otimes{\cal G}_{\nu}^{\bot}$, we shall
restrict our attention to the operator
$H_{\alpha}\upharpoonright\,{\cal H}_{\nu}$; for the sake of
simplicity we shall denote the restriction by the symbol
$H_{\alpha}$ again. The trivial part is absent, of course, if
$\nu$ is irrational.}
\end{rem}

\noindent
From the spectral condition (\ref{spc}) we get a sequence of
eigenvalues (in the natural ascending order) of (the nontrivial part
of) the transverse operator; we denote it as $\{\nu_{n}
(\alpha)\}_{n=1}^{\infty}$. The corresponding eigenfunctions are
\begin{equation}\label{eigenfunctions}
  \chi_{n}(y;\alpha)= (-1)^j N_{n}\, \sin \sqrt{\nu_{n}}d_{j}\, \sin
\sqrt{\nu_{n}}\left(y\!+\!(-1)^jd_{j}\right)
\end{equation}
for $y\in{\cal O}_{j},\; j=1,2$, where $N_{n}$ is the normalization
factor chosen in such a way that $\chi_n$ would be a unit vector in
$\sii(-d_{2},d_{1})$, \ie
\begin{equation}\label{norms}
  N_{n}^{2}=
    \frac{2\sqrt{\nu_{n}}}
    {\sqrt{\nu_{n}}d_{1}\sin^{2}\sqrt{\nu_{n}}d_{2}+
    \sqrt{\nu_{n}}d_{2}\sin^{2}\sqrt{\nu_{n}}d_{1}-
    \sin\sqrt{\nu_{n}}d_{1}\sin\sqrt{\nu_{n}}d_{2}\sin\sqrt{\nu_{n}}D}\,.
\end{equation}
Furthermore, the Green's function of the Hamiltonian (\ref{formal})
can be written down explicitly:
\begin{equation}\label{green0}
  G_{\alpha}(x,y,x',y';k)=\sum_{n=1}^{\infty}\, \frac{i}{2 k_{n}}\,
  e^{i k_{n}|x-x'|}\ \chi_{n}(y;\alpha)\,\bar\chi_{n}(y';\alpha)\,,
\end{equation}
where the efective momentum in the $n$-th transverse mode is
$k_{n}:=\sqrt{k^2-\nu_{n}(\alpha)}$.

Elementary properties of the transverse eigenvalues follow from the
the spectral condition~(\ref{spc}) by means of the implicit-function
theorem; we collect them in the lemma below.
\begin{lemma}\label{approxbound}
(a) Let $\{m_{i}\}_{i=0}^{\infty}$ be the sequence obtained from the
set $\Nat\,\cup\,\nu^{-1}\Nat$ by natural ordering. Then
$ \frac{\pi}{2 d}(n\!-\!1)
  \leq\frac{\pi}{d}m_{n-1}
  <\sqrt{\nu_{n}}
  <\frac{\pi}{d}m_{n}
  \leq\frac{\pi}{d}n\,$
holds for all $n\in\Nat\setminus\{0,1\}$. \\ [1mm]
(b) The function $\alpha\mapsto\nu_{n}(\alpha)$ is strictly
increasing and continuous for all $n\in\Nat\setminus\{0\}$.
\end{lemma}

\setcounter{equation}{0}
\section{Existence of bound states}\label{Sec.Existence}

Depending on the choice of $\alpha$, the operators (\ref{formal})
offer a variety of spectral types. In this paper we shall concentrate
on the situation when the barrier describes a local perturbation of
the system with separating variables considered above. The locality
is at that understood as a decay of the function $\alpha$; in other
words, we shall assume that $\lim_{|x|\to\infty}\alpha(x)=\alpha_0$.
It is important that the limiting value $\alpha_0$ is the same at
both directions.

In such a case, it is easy to localize the essential spectrum. One
employs a simple bracketing argument similar to that
of~\cite[Sec.~II]{ESTV}) squeezing $H_{\alpha}$ between a pair of
operators with Dirichlet and Neumann conditions on segments
perpendicular to the $x$-axis placed to both sides of the centre.
By the minimax principle only the tails of the estimating operators
contribute to their essential spectra; since the ``cuts" can be
chosen arbitrarily far we obtain $\sigma_{ess}(H_{\alpha})=
[\nu_{1}(\alpha_{0}),\infty)$.

Less trivial is the existence of discrete spectrum. It is known
that any ``window" in the impenetrable barrier induces a bound
state. This fact was established first for sufficiently wide
windows \cite{Po}, later an independent and more general proof
proof was given \cite{ESTV} with no lower bound on the window
width. The present case is more complicated because the effective
coupling strength $\alpha-\alpha_0$ can be sign--changing. We
shall show that it is sufficient if it is negative in the mean
creating thus a locally average stronger tunelling between the two
channels:
\begin{thm}\label{existence}
  Assume that {\em (i)} $\;\alpha\!-\!\alpha_{0}\in L_{loc}^1
  (\Real)$, {\em (ii)} $\;\alpha(x)\!-\!\alpha_{0}={\cal
  O}(|x|^{-1-\varepsilon})$ for some $\varepsilon>0$ as
$|x|\to\infty$. If $\;\int_{\Real}(\alpha(x)\!-\!\alpha_{0})dx<0$,
then $H_{\alpha}$ has at least one isolated eigenvalue below its
essential spectrum.
\end{thm}
\PF We use a variational argument whose idea comes back to
~\cite{GJ}; see also~\cite{DE1, RB}, and~\cite[Sec.~III]{ESTV})
for a coupled waveguide system. First of all, the assumption (ii)
tells us that $\lim_{|x|\to\infty}|x|^{1+\varepsilon}
(\alpha(x)\!-\!\alpha_{0})=0$, \ie, to any $\delta>0$ there is
$a_{\delta}>1$ such that
\begin{equation}\label{delta}
  |x|>a_{\delta} \;\Rightarrow\; |\alpha(x)\!-\!\alpha_{0}|
  <\frac{\delta}{|x|^{1+\varepsilon}}.
\end{equation}
 It is useful to introduce a shifted energy form: for an arbitrary
 $\Psi\in D(q_{\alpha})$ we put
\begin{equation}\label{Q}
  Q_{\alpha}[\Psi]:=q_{\alpha}[\Psi]-\nu_{1}(\alpha_{0})\|\Psi\|_{2}^{2};
\end{equation}
since the essential spectrum of $H_{\alpha}$ starts at
$\nu_{1}(\alpha_{0})$, we have to find a trial function $\Psi$
such that $Q_{\alpha}[\Psi]$ is negative. We obtain it by a
suitable modification of the function $\Psi_0(x,y):=
\chi_{1}(y;\alpha_{0})$ which formally annuls (\ref{Q}) for
$\alpha=\alpha_0$ but does not belong to $L^2$. The trial function
has to decay; in order to make the positive contribution from its
tails to the kinetic energy small, we employ an exterior scaling.
We choose an interval ${\cal A}:=[-a,a]$ for some $a>1$ and a
function $\varphi\in{\cal S}(\Real)$ in such a way that
$\varphi(x)\le 1$ and $\varphi(x)=1$ on ${\cal A}$. Then we can
define the family $\{\varphi_{\sigma}:\sigma\in\Real\}$ by a
scaling exterior to ${\cal A}$:
\begin{equation}\label{scaling}
  \varphi_{\sigma}(x):=\left\{
\begin{array}{lcl}
  \varphi(x)                    &\quad\mbox{if}\quad& |x|\leq a\\
  &&\\
  \varphi(\pm a+\sigma(x\mp a)) &\quad\mbox{if}\quad& \pm x>a
\end{array}
\right.
\end{equation}
By construction, $|\varphi_{\sigma}(x)|\leq 1$ holds for all
$x\in\Real$. The sought trial function will be chosen in the form
$\Psi(x,y):= \varphi_{\sigma}(x)\chi_{1}(y;\alpha_{0})$. We employ
the relations $\|\dot{\varphi}_{\sigma}\|_{2}^{2}=
\sigma\|\dot{\varphi}\|_{2}^{2}$, and
\begin{eqnarray*}
  q_{\alpha}[\Psi]&=&q_{\alpha_{0}}[\Psi]+
  \int_{\Real}(\alpha(x)\!-\!\alpha_{0})|\Psi(x,0)|^2 dx,\\
  q_{\alpha_{0}}[\Psi]&=&\|\dot{\varphi}_{\sigma}\|_{2}^{2}
  +\nu_{1}(\alpha_{0})\|\varphi_{\sigma}\|_{2}^{2}\,,
\end{eqnarray*}
the last one of which is obtained by tedious but straightforward
calculation. This yields
\begin{equation}\label{Q=}
  Q_{\alpha}[\Psi]=\sigma\|\dot{\varphi}\|_{2}^{2}
  +|\chi_{1}(0;\alpha_{0})|^2\int_{\Real}
   (\alpha(x)\!-\!\alpha_{0})|\varphi_{\sigma}(x)|^2 dx\,.
\end{equation}
We split now the integration region into two mutually disjoint
parts, ${\cal A}$ and $\Real\setminus{\cal A}$. Using
(\ref{delta}) together with the above mentioned bound on
$\varphi_{\sigma}$ we arrive at the estimate
\begin{equation}\label{Q<}
Q_{\alpha}[\Psi]<\sigma\|\dot{\varphi}\|_{2}^{2}
+\frac{4\,\delta\,|\chi_{1}(0;\alpha_{0})|^2}{\varepsilon
a^{\varepsilon}}
+|\chi_{1}(0;\alpha_{0})|^2\int_{\Real}(\alpha(x)\!-\!\alpha_{0})\,dx\,.
\end{equation}
By assumption we have $\int_{\Real}(\alpha(x)\!-\!\alpha_{0})
\,dx<0$ and since $\chi_{1}(0;\alpha_{0})$ is nonzero, the last
term is negative; it is then enough to choose $\delta$ and
$\sigma$ sufficiently small to make $Q_{\alpha}[\Psi]$
negative.\qed
\begin{rem}\label{WeakRem}{\rm
A case of particular interest concerns weakly coupled Hamiltonian
of the type (\ref{formal}), \ie\/ the situation when $\alpha$
differs from $\alpha_0$ only slightly. In that case one can
develop a Birman-Schwinger analysis in order to derive the
perturbative expansion of the ground state energy in terms of a
parameter measuring the ``smallness" of $\alpha\!-\!\alpha_0$.
This will be done in a separate paper \cite{EK}; here we just
borrow a result for a further use in this work.

There are different ways in which $\alpha\!-\!\alpha_0$ can be
small. Suppose that the support of the perturbation shrinks, \ie\/
introduce $\alpha_{\sigma}(x):= \alpha(x/\sigma)$ with the scaling
parameter $\sigma\in(0,1]$ and consider the limit $\sigma\to 0+$.
We have the following result~\cite{EK}:
\begin{thm}\label{WeakThm}
Suppose that $\alpha\!-\!\alpha_{0}$ is non-zero and belongs to
$L^{1+\varepsilon}(\Real,dx) \cap\si(\Real,|x|^2 dx)$ for some
$\varepsilon>0$. Then $H_{\alpha_{\sigma}}$ has for small $\sigma$
at most one simple eigenvalue $E(\sigma)<\nu_{1}(\alpha_0)$, and
this happens if and only if
$\int_{\Real}(\alpha(x)\!-\!\alpha_{0})\,dx\leq 0$. If this
condition holds the following expansion is valid
\begin{eqnarray}\label{expansion}
\lefteqn{\sqrt{\nu_{1}-E(\sigma)} =
-\frac{\sigma}{2}\,|\chi_{1}(0;\alpha_0)|^2
  \int_{\Real}(\alpha(x)\!-\!\alpha_{0})\,dx} \nonumber\\
&&
  +{\sigma^2\over 4}\,
  |\chi_{1}(0;\alpha_0)|^2\sum_{n=2}^{\infty}|\chi_{n}(0;\alpha_0)|^2
  \int_{{\Real}^2}(\alpha(x)\!-\!\alpha_{0})\,
  \frac{e^{-\sigma\sqrt{\nu_{n}-\nu_{1}}|x-x'|}}
  {\sqrt{\nu_{n}-\nu_{1}}}\,
  (\alpha(x')\!-\!\alpha_{0})\,dx\,dx\nonumber\\
&&
  +{\cal O}(\sigma^3).
\end{eqnarray}
\end{thm}}
\end{rem}

\setcounter{equation}{0}
\section{A ``rectangular well'' example}\label{Sec.Example}
To illustrate the above result and to analyze the behaviour of
coupled waveguides in more details we shall now investigate an
example. We choose the barrier function $\alpha$ so that the
corresponding Schr\"odinger equation can be solved numerically;
this happens if $\alpha$ is a step-like function which makes it
possible to employ the mode-matching method. The simplest
nontrivial case concerns a ``rectangular well" of a width $2a>0$,
\begin{displaymath}
  \alpha(x):=\left\{
  \begin{array}{lcl}
    \alpha_{1} & \mbox{if} & |x|<a\\
    &&\\
    \alpha_{0} & \mbox{if} & |x|\leq a
  \end{array}
  \right.
\end{displaymath}
for some $\alpha_{1},\alpha_{0}\in\Real$. In view of
Theorem~\ref{existence} this waveguide system has bound states if
and only if $\alpha_1< \alpha_0$. In particular, one expects that
in the case when $\alpha_{1}=0$ and $\alpha_{0}$ is large positive
the spectral properties will be similar to those of the
impenetrable barrier situation studied in~\cite{ESTV}. On the
other hand, the mode-matching method allows us to treat on same
footing the scattering processes in our waveguide. Then there is
no need to impose the above condition, because the ``barrier"
situation, $\alpha_1> \alpha_0$ is expected to exhibit nontrivial
scattering behaviour as well.

Henceforth, we shall denote the transverse eigenvalues in the two
regions as $\nu_n^{s}:=\nu_n(\alpha_s)$, $s:=0,1$,
$\;n\in\Nat\setminus\{0\}$. In view of the natural mirror symmetry
with respect to the $y$-axis we may consider separately the
symmetric and antisymmetric solutions, \ie\/ to analyze the
halfstrip with the Neumann or Dirichlet boundary condition at
$x=0$, respectively. For the sake of simplicity we shall also
restrict our attention to the case $\;\min\{\alpha_0,\alpha_1\}
>\alpha_m:= -(d_{1}^{-1}+d_{2}^{-1})$, when the lowest transverse
eigenvalue is positive everywhere in the waveguide. The
considerations presented below remain valid even without this
assumption; one has just to replace the trigonometric ground-state
eigenfunction for hyperbolic which makes the formulae cumbersome.

\subsection{Bound states} \label{Sec.existence}
Let us first derive an estimate which allows to localize roughly the
eigenvalues. It is based on a bracketing argument similar to that
used to specify the essential spectrum at the beginning of
Section~\ref{Sec.Existence}. The Hamiltonian can be squeezed between
a pair of operators, $H_{\alpha}^{(N)}\leq H_{\alpha}\leq
H_{\alpha}^{(D)}$, with additional Dirichlet/Neumann ``cuts" at
segments perpendicular to the waveguide axis, $x=\pm a$. The spectra
of the estimating operators can be easily found and sought estimate
comes from the eigenvalues of the middle part situated below
$\nu^0_1$ in combination with the minimax principle. In particular,
we find that the number $N$ of isolated eigenvalues satisfies the
bounds
$$
 N_{D}+1\geq N\geq
 N_{D}:=\left[\frac{2 a}{\pi}\sqrt{\nu_{1}^{0}-\nu_{1}^{1}}\right]\,,
$$
where $[\cdot]$ denotes the entire part; this complements
Theorem~\ref{existence}. Furthermore, the $n$-th eigenvalue
$E_{n}$ of $H_{\alpha}$ is estimated by
\begin{equation}\label{estE}
  \nu_{1}^{1}+\left(\frac{(n-1)\pi}{2 a}\right)^2
  \leq E_n \leq
  \nu_{1}^{1}+\left(\frac{n\pi}{2 a}\right)^2,
\end{equation}
while the critical halfwidth $a_n$ at which the $n$-th eigenvalue
emerges from the continuum satisfies the bounds
\begin{equation}\label{estg}
  \frac{(n-1)\pi}{2\sqrt{\nu_1^{0}-\nu_1^{1}}}
  \leq a_n \leq
  \frac{n\pi}{2\sqrt{\nu_1^{0}-\nu_1^{1}}}.
\end{equation}
After this preliminary, let us pass to the mode-matching method. We
start with the simpler case when the waveguide exhibits a mirror
symmetry w.r.t. the $x$-axis, \ie\/ $d_1=d_2=d$.

\subsubsection{The symmetric case}
If $\nu=1$, the Hamiltonian decouples into an orthogonal sum of the
even and the odd parts, the spectrum of the latter being clearly
trivial --- \cf~Remark~\ref{rem1}. The two symmetries allow us to
restrict ourselves to the part of $\Omega$ in the first quadrant,
with Neumann or Dirichlet condition in the segment $(0,d)$ of the
$y$-axis, and take the transverse eigenvalues determined by the
spectral conditions
\begin{eqnarray*}
  -\alpha_0   &=& 2 \ell\cot\ell d \qquad\mbox{if}\quad x\geq a\\
  -\alpha_1   &=& 2 \ell\cot\ell d \qquad\mbox{if}\quad 0\leq x<a.
\end{eqnarray*}
The corresponding transverse eigenfunctions are
\begin{eqnarray}
  \chi_{n} &:=& -N_n^{0}\sin\sqrt{\nu_n^{0}}(y\!-\!d)
  \qquad\mbox{if}\quad x\geq a,\nonumber \\
  \phi_{n} &:=& -N_n^{1}\sin\sqrt{\nu_n^{1}}(y\!-\!d)
  \qquad\mbox{if}\quad 0\leq x<a,
\end{eqnarray}
where $N_n^{s}$ is a normalization factor chosen to make
$\chi_n,\phi_n$ unit vectors in $\sii(0,d)$, \ie
\begin{equation}
  (N_n^{s})^2=\frac{4\sqrt{\nu_n^{s}}}
  {2\sqrt{\nu_n^{s}}d-\sin 2\sqrt{\nu_n^{s}}d}\ .
\end{equation}
The overlap integrals of elements of the two bases are easily seen
to be
\begin{equation}\label{overlap}
  (\chi_m,\phi_n)=\frac{N_m^0 N_n^1}{\nu_m^0-\nu_n^1}
  \left(\sqrt{\nu_n^1}\sin\sqrt{\nu_m^0}d\,\cos\sqrt{\nu_n^1}d
  -\sqrt{\nu_m^0}\sin\sqrt{\nu_n^1}d\,cos\sqrt{\nu_m^0}\right).
\end{equation}
A natural Ansatz for the solution of an energy
$E\in[\nu_1^1,\nu_1^0)$ is
\begin{equation}\label{Ansatz}
\begin{array}{rll}
  \Psi_{s/a}(x,y)&=\sum\limits_{n=1}^{\infty}b_n^{s/a} e^{-q_n(x-a)}\chi_n(y)
  &\qquad\mbox{for}\quad x\geq a\\
  &&\\
  \Psi_{s/a}(x,y)&=\sum\limits_{n=1}^{\infty}a_n^{s/a}
  \left\{
  \begin{array}{c}
    \frac{\cosh p_n x}{\cosh p_n a}\\ \\
    \frac{\sinh p_n x}{\sinh p_n a}
  \end{array}
  \right\}
  \phi_n(y)
  &\qquad\mbox{for}\quad 0\leq x<a
\end{array}
\end{equation}
where the subscripts (we will omit them for the most part) $s,a$
distinguish the symmetric and antisymmetric case, respectively.
The longitudinal momenta are defined by
\begin{displaymath}
  q_n:=\sqrt{\nu_n^0-E}\ ,\qquad p_n:=\sqrt{\nu_n^1-E}\ .
\end{displaymath}
As an element of the domain~(\ref{Hamiltonian}), the function
$\Psi$ should be continuous together with its normal derivative at
the segment dividing the two regions, $x=a$. Using the
orthonormality of $\{\chi_n\}$ we get from the requirement of
continuity
\begin{equation}\label{contRequir}
  b_m=\sum_{n=1}^{\infty}a_n(\chi_m,\phi_n)\,.
\end{equation}
In the same way, the normal-derivative continuity at $x=a$ yields
\begin{equation}\label{normal-derRequir}
  b_m q_m+\sum_{n=1}^{\infty}a_n p_n
  \left\{
  \begin{array}{c}
    \tanh\\
    \coth
  \end{array}
  \right\}
  (p_n a) (\chi_m,\phi_n)=0\,.
\end{equation}
Substituting from~(\ref{contRequir}) to~(\ref{normal-derRequir}),
we can rewrite it as an operator equation
\begin{equation}\label{C.a=0}
  \boldmath{C a}=\boldmath{0}\,,
\end{equation}
where
\begin{equation}
  C_{mn}:=\left(q_m+p_n
  \left\{
  \begin{array}{c}
    \tanh\\
    \coth
  \end{array}
  \right\}
  (p_n a)\right)(\chi_m,\phi_n)
\end{equation}
with the overlap integrals given by~(\ref{overlap}).

    It is straightforward to compute the norms of
the functions~(\ref{Ansatz}); since $n^{-1}q_n$ an $n^{-1}p_n$
tend to $\frac{\pi}{d}$  as $n\to\infty$
(see~Lemma~\ref{approxbound}~1.), the square integrability of
$\Psi$ requires the sequences $\{a_n\}$ and $\{b_n\}$ to belong to
the space $\ell^2(n^{-1})$.

    To make sure that the equation~(\ref{C.a=0}) makes sense,
it is enough to notice that if $\Psi$ is an eigenvector of
$H_{\alpha}$, it must belong to the domain of any integer power of
this operator. It is easy to check that
\begin{equation}
  \forall i\in\Nat\!\setminus\!\{0\}:\quad
  \Psi\in D(H_{\alpha}^i)\Leftrightarrow
  \{a_n\},\{b_n\}\in\ell^2(n^{4 i-1})\,;
\end{equation}
hence the sought sequences should belong to $\ell^2(n^r)$ for all
$r\geq -1$, \ie\ both sequences have a faster than powerlike
decay. This fact also justifies {\em a posteriori\/} the interchange
of summation and differentiation we have made in the matching
procedure. Furthermore, one can use it to check the existence of a
convergent series of cut-off approximants to the solutions in the
same way as in~\cite[Sec.~IV.1]{ESTV}.
\begin{rem}[an alternative method]{\rm
We can use the orthonormality of $\{\phi_n\}$ instead of
$\{\chi_n\}$ and express $\{a_n\}$ in analogy
to~(\ref{contRequir}), and then substitute it
into~(\ref{normal-derRequir}). We find that the coefficient
sequence $\{b_n\}$ is then determined by the following equation:
\begin{equation}\label{b+K.b=0}
  \boldmath{b}+\boldmath{K b}=\boldmath{0}\,,
\end{equation}
where
\begin{equation}
  K_{mn}:=\frac{1}{q_m}\sum_{k=1}^{\infty}(\chi_m,\phi_k)\,p_k
  \left\{
  \begin{array}{c}
    \tanh\\
    \coth
  \end{array}
  \right\}
  (p_k a)\,(\phi_k,\phi_n)\,.
\end{equation}
The two approaches~(\ref{C.a=0}),~(\ref{b+K.b=0}) are, of course,
equivalent, however, it may be useful to combine them in order to get 
a good idea about the numerical stability of the solution.
For instance, in the situation of~\cite{ESTV} the approximants
of~(\ref{C.a=0}) approach the limiting values from above, while
those referring to~(\ref{b+K.b=0}) are increasing.}
\end{rem}

\subsubsection{The Asymmetric Case}
    Let us pass now to the case, when the widths of the ducts
are nonequal, $\nu\not=1$. In view of the mirror symmetry, we
shall consider right-halfplane part of $\Omega$ only with  the
Neumann and Dirichlet condition on the segment $[-d_2,d_1]$ of the
$y$-axis. The asymmetric case differs from the previous one just
in the choice of transverse basis: now we can take
\begin{eqnarray}
  \chi_{n}(y) &:=& \chi_n(y;\alpha_0)
  \qquad\mbox{if}\quad x\geq a \nonumber\\
  \phi_{n}(y) &:=& \chi_n(y;\alpha_1)
  \qquad\mbox{if}\quad 0\leq x<a,
\end{eqnarray}
where $\chi_n(\cdot,\alpha_s)$ are of the
form~(\ref{eigenfunctions}) with the norms $N_n^s$ given
by~(\ref{norms}). The corresponding eigenvalues $\nu_n^0, \nu_n^1$
are then determined by
\begin{eqnarray*}
  -\alpha_0 &=& \ell\,(\cot\ell d_1+\cot\ell d_2)
   \qquad\mbox{if}\quad x\geq a\\
  -\alpha_1 &=& \ell\,(\cot\ell d_1+\cot\ell d_2)
   \qquad\mbox{if}\quad 0\leq x<a\,.
\end{eqnarray*}
(see~(\ref{spc})) and the overlap integrals are
\begin{eqnarray}\label{ASoverlap}
  (\chi_m,\phi_n) &=& \frac{N_m^0 N_n^1}{\nu_m^0-\nu_n^1}\
  \Big(\sqrt{\nu_n^1}\,\sin\sqrt{\nu_m^0}d_1\,sin\sqrt{\nu_m^0}d_2
  \,\sin\sqrt{\nu_n^1}D\nonumber\\
      &&\qquad\quad\
  -\sqrt{\nu_m^0}\,\sin\sqrt{\nu_n^1}d_1,\sin\sqrt{\nu_n^1}d_2
   \,\sin\sqrt{\nu_m^0}D\Big)\,.
\end{eqnarray}
The rest of the argument does not change and one has again to
solve the equation~(\ref{C.a=0}) (respectively,~(\ref{b+K.b=0})).
By a straightforward modification of the above argument, one can
also check that the coefficient sequences have a faster than
powerlike decay and that the equation can be solved by a sequence
of truncations.


\begin{figure}[!tb] 
\begin{center}
\includegraphics[width=\textwidth]{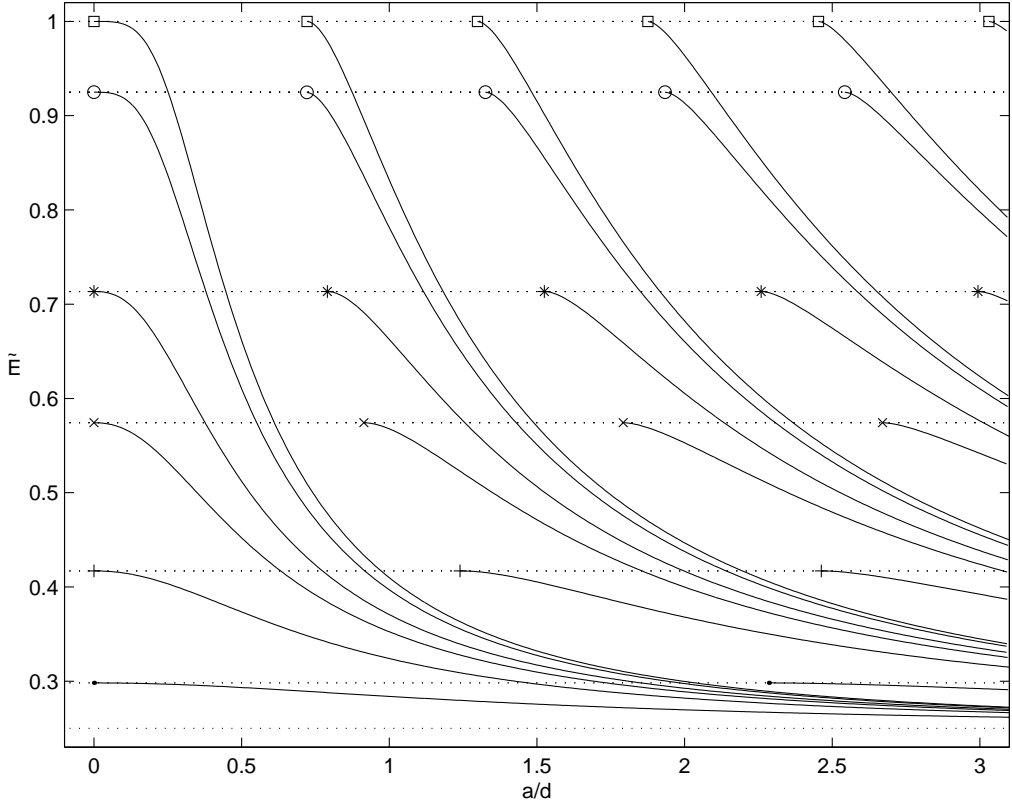}
\end{center}
\caption{Bound state energies {\em vs.} the halfwidth $\tilde a$ 
         in the symmetric case for $\tilde\alpha_0=10^5\ (\square)$,
         $50\ (\circ)$, $10\ (\ast)$, $5\ (\times)$, $2\ (+)$,
         $0.5\ (\bullet)$.}\label{figEigenvalues}
\end{figure}

\begin{figure}[!tb] 
\begin{center}
\includegraphics[width=\textwidth]{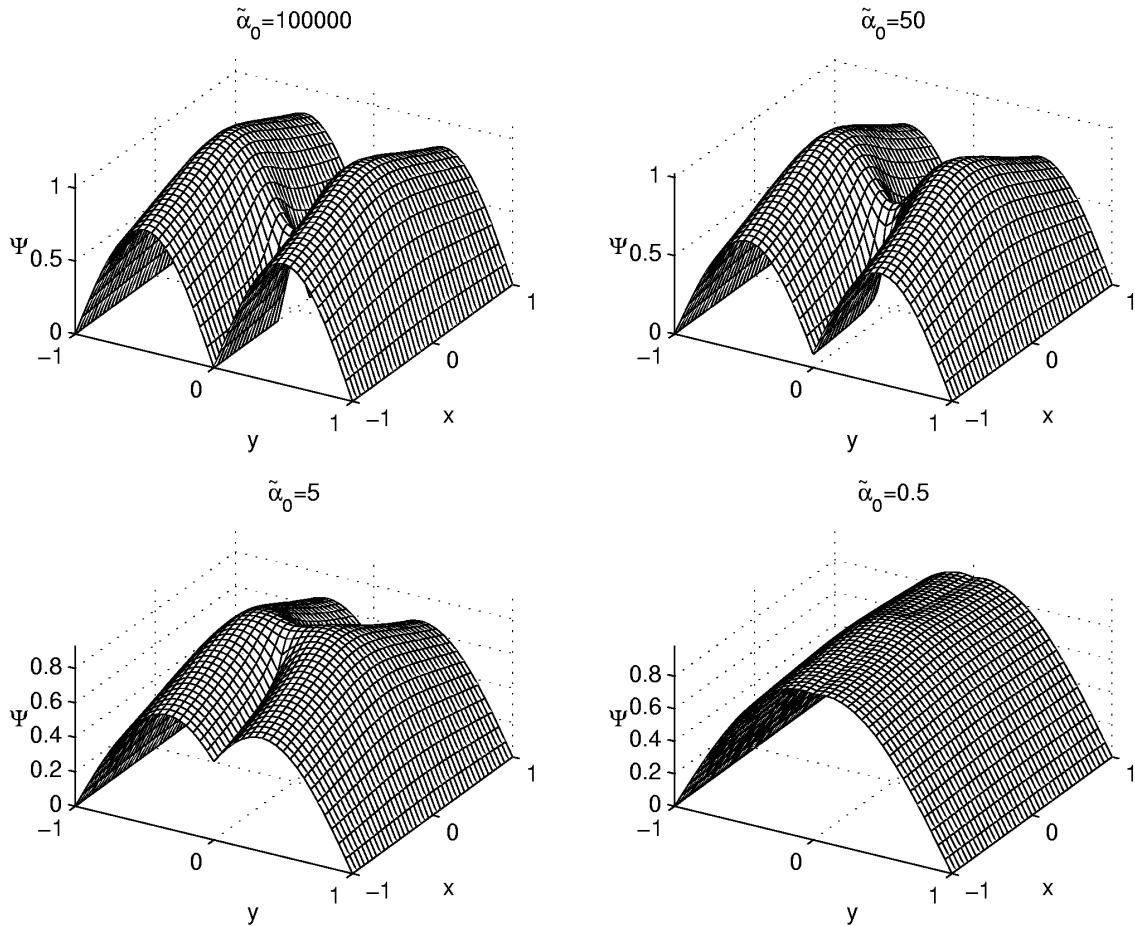}
\end{center}
\caption{Ground-state eigenfunctions in the symmetric case 
         for $a/d=0.15$.}\label{figEigenfunctions}
\end{figure}

\begin{figure}[!tb] 
\begin{center}
\includegraphics[width=\textwidth]{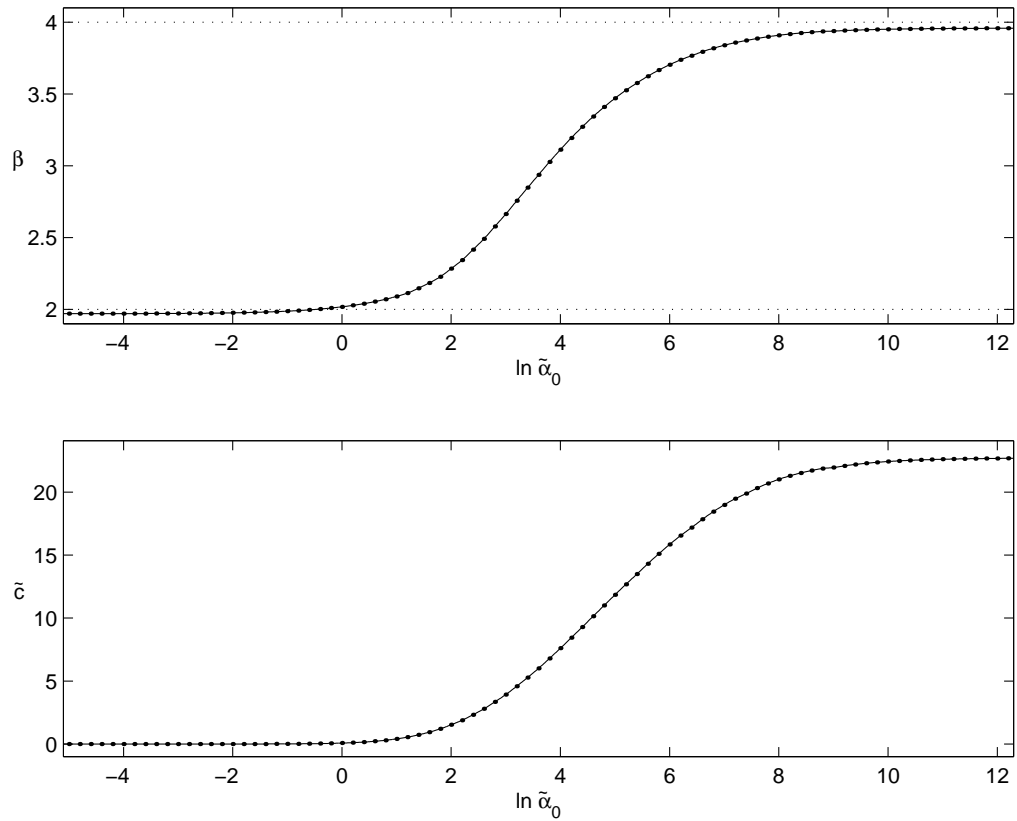}
\caption{Narrow-window asymptotic power and coefficient as
	      functions of $\alpha_0$.}\label{fig24}
\end{center}

\end{figure}

\begin{figure}[!tb] 
\begin{center}
\includegraphics[width=100mm]{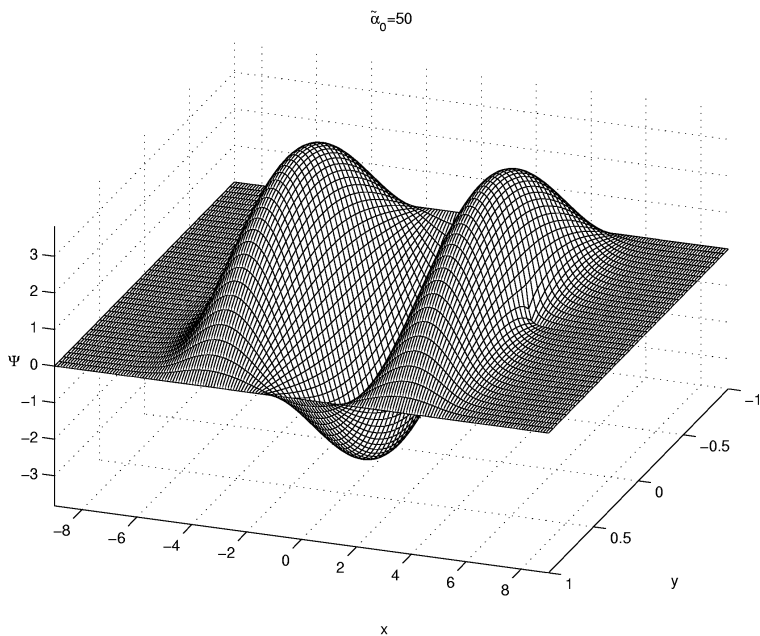}
\caption{The eigenfunction of the second excited state 
			for $\nu=1$, $a/d=5$,
         $\alpha_1=0$.}\label{figExcitedEigenfunction}
\end{center}
\end{figure}

\begin{figure}[!tb] 
\begin{center}
\includegraphics[width=\textwidth]{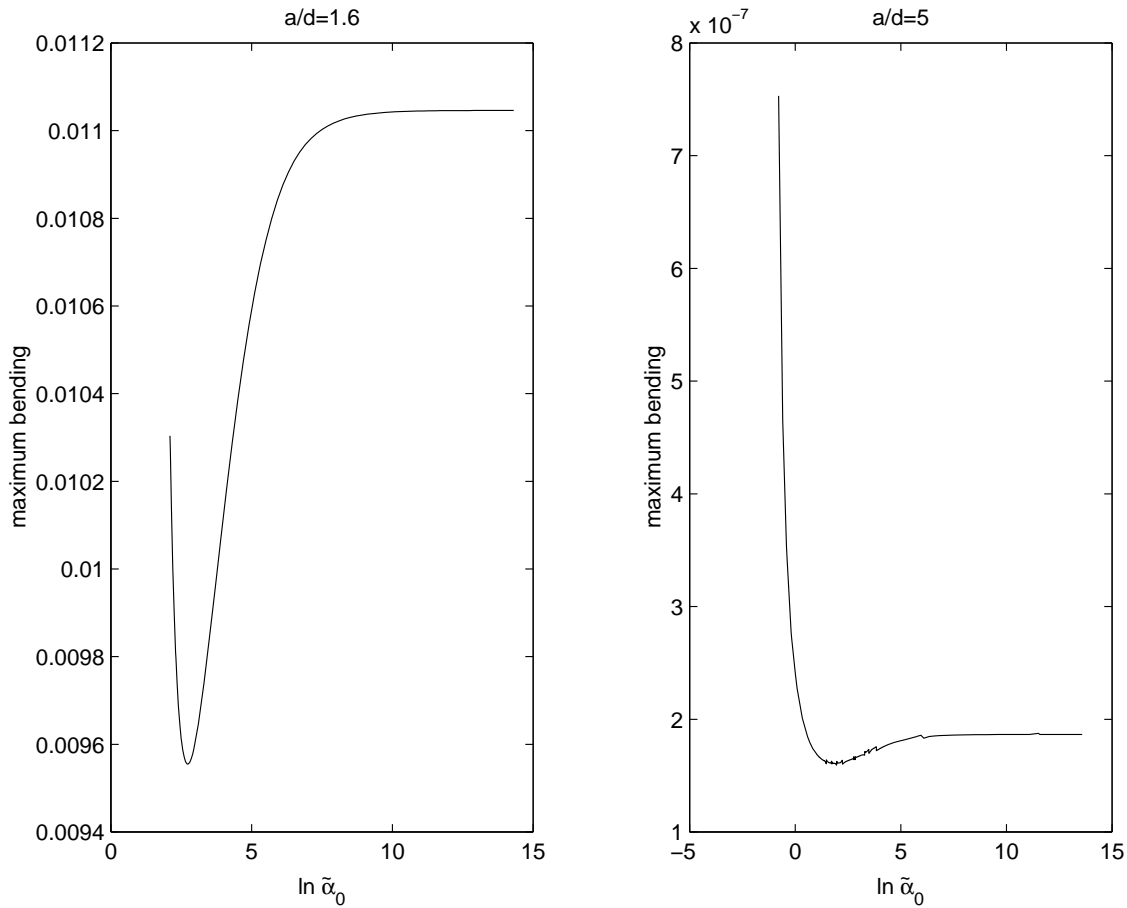}
\caption{The maximum bending of nodal lines of third eigenfunctions  
			in the symmetric case as a function of~$\alpha_0$ for 
  			a fixed window width.}\label{figNodalLines}
\end{center}
\end{figure}

\begin{figure}[!tb] 
\begin{center}
\includegraphics[width=100mm]{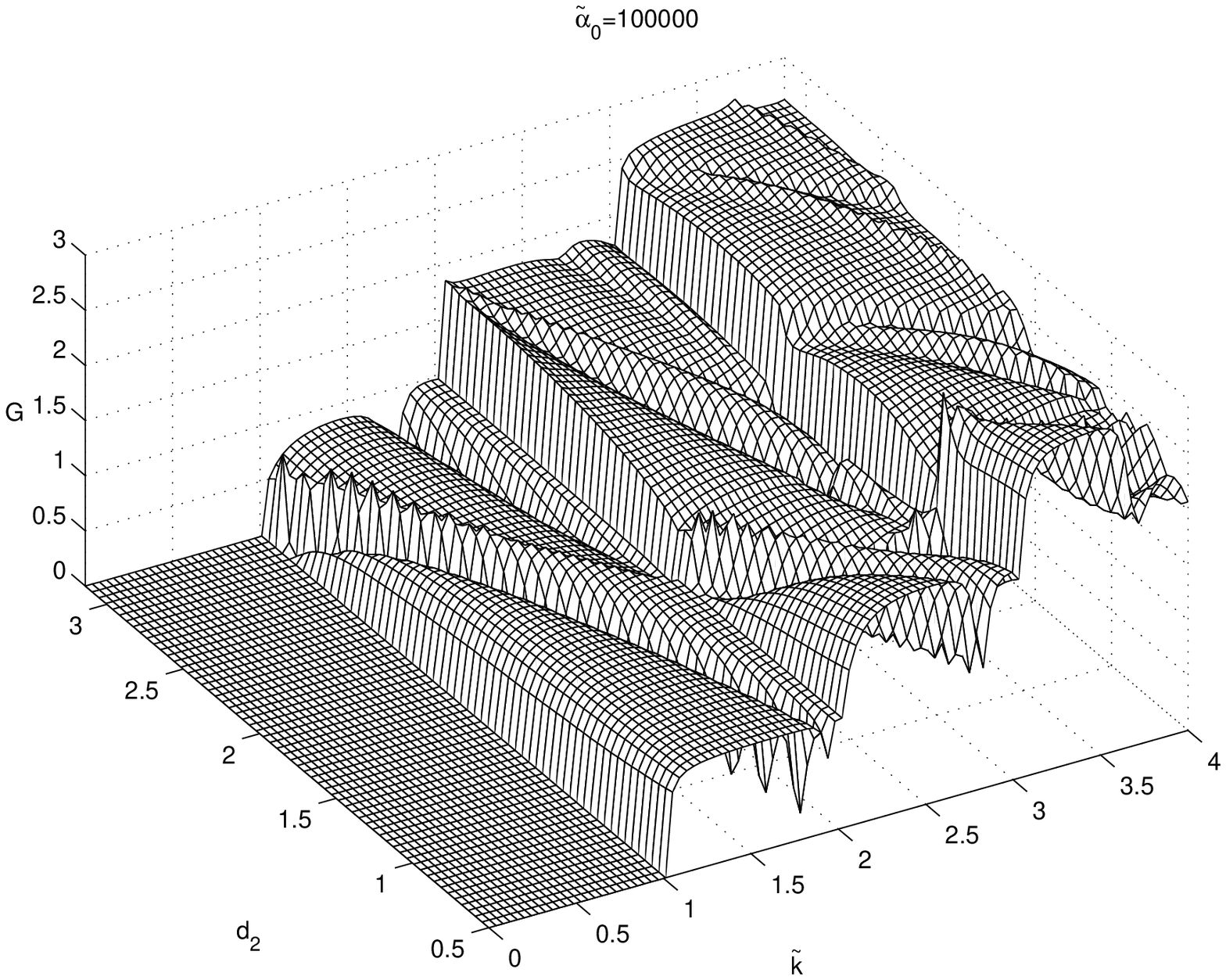}
\includegraphics[width=100mm]{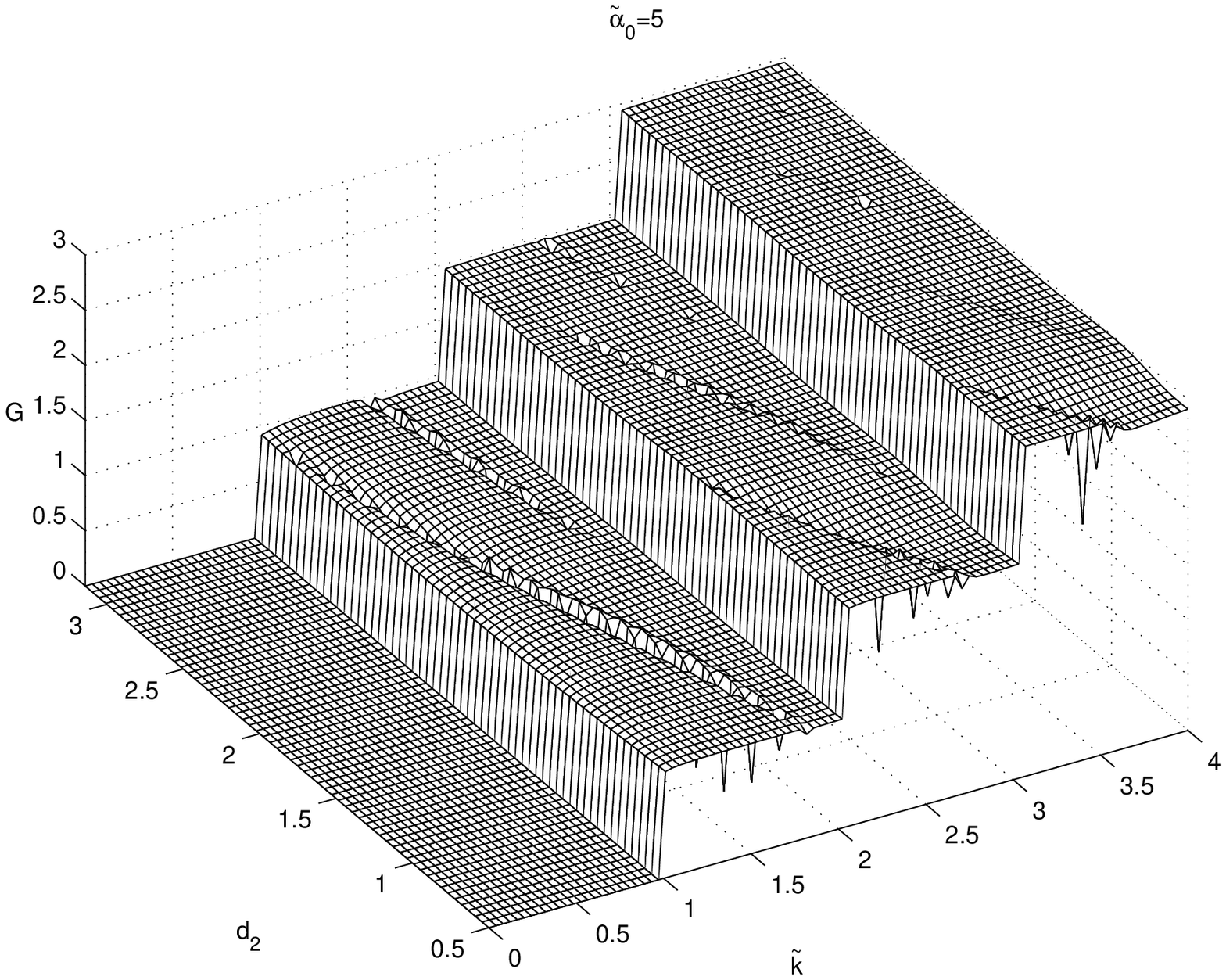}
\caption{Right-left conductivity as a function of $k$, $d_2$
			for $d=\pi$, $a=2\pi$, $\alpha_1=0$.}\label{figConductivity}
\end{center}
\end{figure}

\begin{figure}[!tb]
\begin{center}
\includegraphics[width=\textwidth]{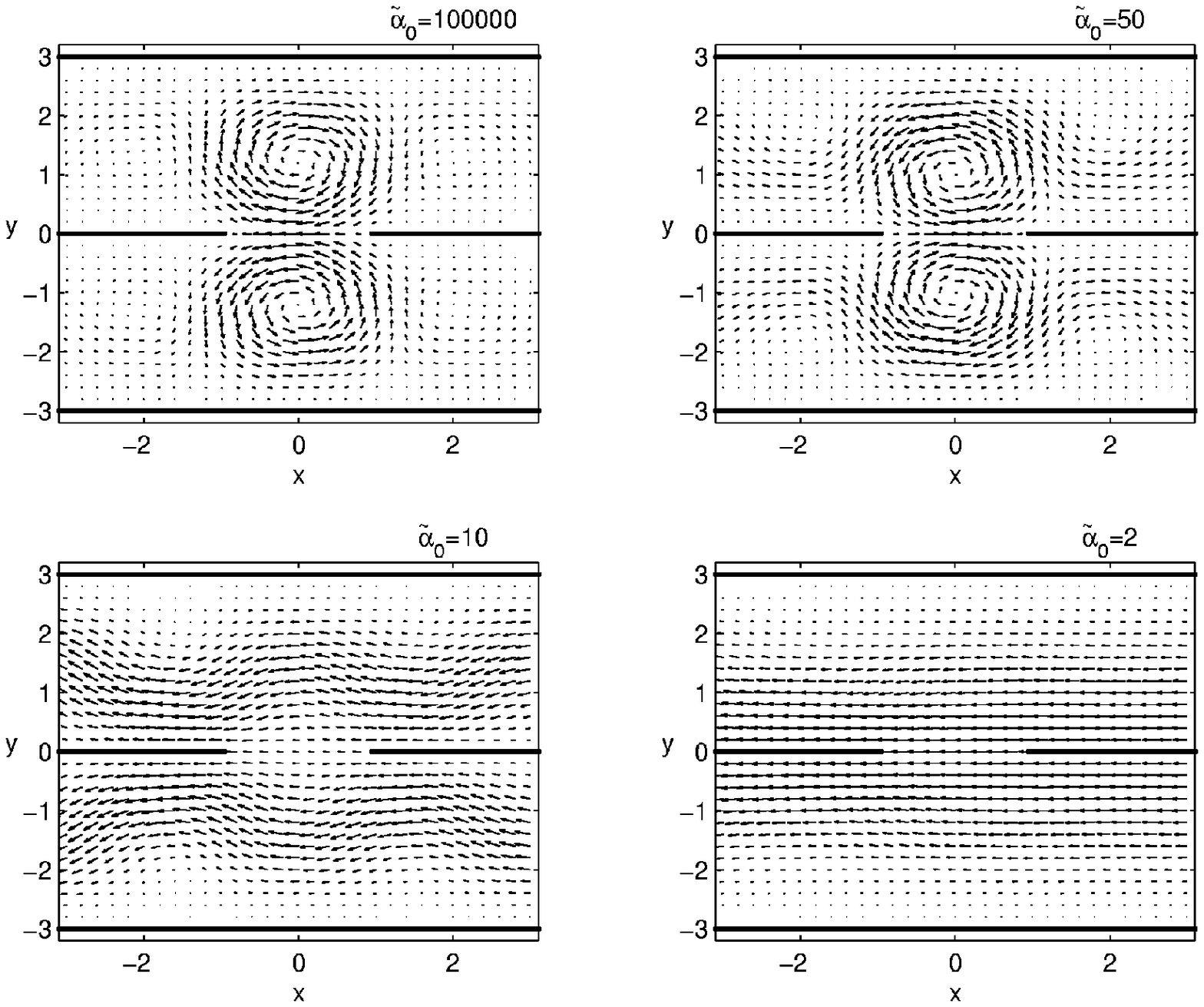}
\caption{Probability flow patterns for $\tilde k=1.745$ and $\alpha_1=0$ in 
         the symmetric situation.}\label{figFlow}
\end{center}
\end{figure}

\subsection{Scattering}
    As we said in the opening of this section, the scattering can
be treated in an analogous way. The incident wave is supposed to
be of the form $e^{-i k_r x} \chi_r(y;\alpha_0)$, \ie, to come
from the right in the $r$-th transverse mode; we have introduced
the effective momentum $k_r:=\sqrt{k^2-\nu_r^0}$. We denote by
$r_{rn}, t_{rn}$, respectively, the corresponding reflection and
transmission amplitudes to the $n$-th transverse mode. Due to the
mirror symmetry, we can again separate the symmetric and
antisymmetric situation w.r.t. $x=0$ and to write
\begin{equation}\label{amplitudes}
  r_{rn}=\frac{1}{2}(\rho_{rn}^s+\rho_{rn}^a)\,,\qquad
  t_{rn}=\frac{1}{2}(\rho_{rn}^s-\rho_{rn}^a)\,,
\end{equation}
where $\rho_{rn}^{\sigma}$, $\sigma=s,a$, are the reflection
amplitudes in a half of our waveguide with the Neumann and
Dirichlet condition at $x=0$, respectively. We use the following
Ansatz
\begin{equation}
\begin{array}{rll}
  \Psi_{s/a}(x,y)&=\sum\limits_{n=1}^{\infty}
  \left(\delta_{rn}e^{-i k_n (x-a)}
  +\rho_{rn}^{s/a}e^{i k_n (x-a)}\right)\chi_n(y)
  &\quad\mbox{for}\quad x\geq a\\
  &&\\
  \Psi_{s/a}(x,y)&=\sum\limits_{n=1}^{\infty}a_n^{s/a}
  \left\{
  \begin{array}{c}
    \frac{\cos p_n x}{\cos p_n a}\\ \\
    \frac{\sin p_n x}{\sin p_n a}
  \end{array}
  \right\}
  \phi_n(y)
  &\quad\mbox{for}\quad 0\leq x<a
\end{array}
\end{equation}
for the total energy $k^2$ of the incident wave in the $r$-th
mode. The quantities $p_n:=\sqrt{k^2-\nu_n^1}$ are effective
momenta in the ``interaction" region. Matching these functions
smoothly at $x=a$ we arrive in the same way as above at the
equation
\begin{equation}\label{C.a=f}
  \boldmath{C a}=\boldmath{f}\,,
\end{equation}
where
\begin{eqnarray}
  C_{mn} &:=& \left(i k_m+p_n
  \left\{
  \begin{array}{c}
    \tan\\
    -\cot
  \end{array}
  \right\}
  (p_n a)\right)(\chi_m,\phi_n)\\
  f_m    &:=& 2 i k_m \delta_{rm},
\end{eqnarray}
where the index $r$ corresponds to the incident wave and the
overlap integrals are given again by~(\ref{ASoverlap}). The
reflection amplitudes are then given by
\begin{equation}
  \rho_{rm}=-\delta_{rm}+\sum_{n=1}^{\infty} a_n (\chi_m,\phi_n)\,;
\end{equation}
they determine the full S-matrix via~(\ref{amplitudes}). A
quantity of direct physical interest is rather the conductivity
given by the Landauer formula. If we express it in the standard
units $2 e^2/h$, it equals
\begin{equation}
  G(k)=\sum_{m,n=1}^{[k]}\frac{k_n}{k_m}|t_{mn}(k)|^2\,,
\end{equation}
where $t_{mn}(k)$ are the coefficients~(\ref{amplitudes}). The
summation runs over all open channels. Another physically
interesting quantity is the probability flow distribution
associated with the generalized eigenvector $\Psi=\Psi_s+\Psi_a$,
which is defined in the standard way,
\begin{equation}
  \vec j(\vec x)=2\im\left(\bar\Psi(\vec x)\nabla\Psi(\vec x)\right).
\end{equation}

\subsection{Numerical results}
Since the spectrum behaves naturally at scaling transformations it
is reasonable to solve the equations~(\ref{C.a=0})
and~(\ref{C.a=f}) in the natural non-dimensional quantities. We
mark them by tilde and use them to label the axis in the figures,
\eg, $a=d\tilde a$, $\alpha_s=\tilde\alpha_s/d$, $E=(\pi/d)^2\tilde E$ 
or $k=(\pi/d)\tilde k$.

\subsubsection{Bound states}
\paragraph{Eigenvalues:}
    Figure~\ref{figEigenvalues} shows the bound-state energies
as functions of the ``window'' halfwidth $a$ for an ``empty
window", $\alpha_1=0$. Several curves referring to different
values of the barrier coupling constant $\alpha_0$ are plotted. In
accordance with the general results of Section~\ref{Sec.existence}
the energies decrease monotonously with the increasing ``window''
width and one can sandwich them between the
estimates~(\ref{estE}). We also see that for a fixed $a$ the
energies increase with respect to $\alpha_0$ and $\nu$; recall
that their number increases as a function of $\alpha_0$ but it
decreases as the waveguide becomes more asymmetric --- these facts
are clear from~(\ref{estE}),~(\ref{estg}), and
Lemma~\ref{approxbound}. It is illustrative to confront our
results for large $\alpha_0$ with the energies computed
in~\cite{ESTV} for the case which corresponds formally to
$\alpha_0=\infty$. Comparing Figure~\ref{figEigenvalues} with
Fig.~2 of the mentioned paper we see that our result for
$\tilde\alpha_0=10^5$ is practically identical with the latter.

\paragraph{Eigenfunctions:}
    The evolution of the ground-state wavefunction
with respect to $\alpha_0$ for an empty window of the fixed
halfwidth $\tilde a=0.15$ is illustrated on
Figure~\ref{figEigenfunctions}. If the barrier tunneling is
negligible, $\tilde\alpha_0=10^5$, the picture is
indistinguishable from Fig.~3 in~\cite{ESTV}. As $\alpha_0$
becomes smaller we see how the wavefunction part penetrating the
barrier grows.

\paragraph{Threshold behaviour:}
Consider again the empty--window case, $\alpha_1=0$. As a
consequence of Theorem~\ref{WeakThm} we get for the ground-state
energy for any fixed $\alpha_0$ and a narrow window the asymptotic
formula (\cf~(\ref{expansion}))
\begin{equation}\label{tresh}
  E(a)=\nu_{1}^0-c\,a^2+{\cal O}(a^3)\ ,\ \qquad
  c:=a^2 \alpha_{0}^2\ |\chi_{1}(0;\alpha_0)|^4.
\end{equation}
On the other hand, in the case of a window in the Dirichlet
barrier, $\alpha_0=\infty$, it was conjectured in~\cite{ESTV} that
we may suppose
\begin{equation}\label{windowResult}
  E(a)=\left(\frac{\pi}{d}\right)^2-C(\nu)\,a^4+{\cal O}(a^5)
\end{equation}
as $a\to 0+$. The conjecture is supported by a two-sided
asymptotic estimate~\cite{EV2}: there are positive $c_{1},c_{2}$
such that
\begin{displaymath}
  -c_{1} a^4\leq E(a)-\left(\frac{\pi}{d}\right)^2\leq-c_{2} a^4
\end{displaymath}
(for a generalization of this result to a larger number of windows
and higher dimensions see~\cite{EV3}). Quite recently, a proof of
(\ref{windowResult}) has been proposed by Popov~\cite{Pop}.

This seems to be a paradox. In order to make sense of these
considerations, we suppose that $E(a)=\nu_{1}^0-c a^{\beta}$ for
small $a$ of an interval $0.016<\tilde a<\tilde a_{max}$
($\tilde a_{max}=\tilde a_{max}(\alpha_0)$ 
is chosen in such a way to include the best correlated points),
and investigate numerically the dependence of the coefficients $\beta$ 
and $c$ ($\tilde c=d^{\beta+2}c/\pi^2$) on
$\alpha_0$. The powerlike asymptotic  behaviour is confirmed when
we redraw the first eigenvalue curves of
Figure~\ref{figEigenvalues} in the logarithmic scale. The obtained
dependence of the coefficients on $\alpha_0$ in the symmetric case
$\nu=1$, is illustrated on Figure~\ref{fig24}. We see that the
power reaches the values $\beta=2,4$ for small and large
$\alpha_0$, respectively (a slight shift in the
first graph is due the truncation; the convergence becomes very
slow for small $a$). At the same time, the numerically found $c$
for small $\alpha_0$ coincides with that of~(\ref{tresh}).

Of course, the asymptotical behaviour is governed by~(\ref{tresh})
for any finite $\alpha_0$. The above result says only that the
transition from biquadratic to quadratic asymptotics occurs for
large $\alpha_0$ at values of $a$ still smaller than those we have
used.

\paragraph{Nodal Lines:}
In Figure~\ref{figExcitedEigenfunction} we plot the third eigenfunction.
Its nodal lines are almost straight
showing thus that the ``spikes'' at the window edges act almost as
a hard barrier. On the other hand, a simple argument based 
on the reflection principle shows that the nodal lines cannot be straight.
Closer inspection shows that they have the form of a bow bent outward.
The maximum bending is shown on Figure~\ref{figNodalLines}.
It decreases rapidly with the window width
which confirms the tunneling nature of the effect. Nodal lines of higher
eigenfunctions exhibit (as functions of~$\alpha_0$) irregularities
connected with the changes in the number of the modes.

\subsubsection{Scattering}
\paragraph{Conductivity:}
    Figure~\ref{figConductivity} illustrates
the evolution of the conductivity for the particle coming from the
right and leaving to the left as a function of the momentum $k$
and the width $d_2$. We see that the perturbation, $-\alpha_0$ in
the window, deforms the ideal steplike shape with jumps at
transverse thresholds; deep resonances are clearly visible. For an
almost impenetrable barrier, $\tilde\alpha_0=10^5$, we practically
reproduce Fig.~5a of~\cite{ESTV}.

\paragraph{Probability flow:}
    Examples of the quantum probability flow are
shown on Figure~\ref{figFlow}. The flow patterns change with the
momentum of the incident particle and the value of $\alpha_0$.
They exhibit conspicuous vortices at the resonance energies which
correspond to the ``trapped part'' of the wavefunction. An
interesting phenomenon is illustrated on
the first two graphs of Figure~\ref{figFlow}: 
for $\tilde\alpha_0=10^5$ there is a
double vortex (corresponding to the sharp stopping resonance of
Figure~\ref{figConductivity}), right-handed in the upper duct,
while for $\tilde\alpha_0=50$ we get a left-handed vortex. The
conductivity is small in these situations so the waveguide system
is effectively closed for the particle transport. As $\alpha_0$
decreases the conductivity grows and the waveguide opens ---
\cf~Figure~\ref{figFlow} for $\tilde\alpha_0=10,2$.

\setcounter{equation}{0}
\section{An Aharonov-Bohm cylinder}

   In the closing section we want to show now how the preceding
considerations modify for a different geometry: we consider a
nonrelativistic quantum particle living on the surface of an
infinite straight cylinder of a radius $R$, which is threaded by a
homogeneous magnetic field $\vec B$ parallel to the cylinder axis.
We assume that the motion is further restricted by a
$\delta$-barrier supported by a line parallel to the axis.

\begin{figure}[!ht] 
\begin{center}
\includegraphics[width=\textwidth]{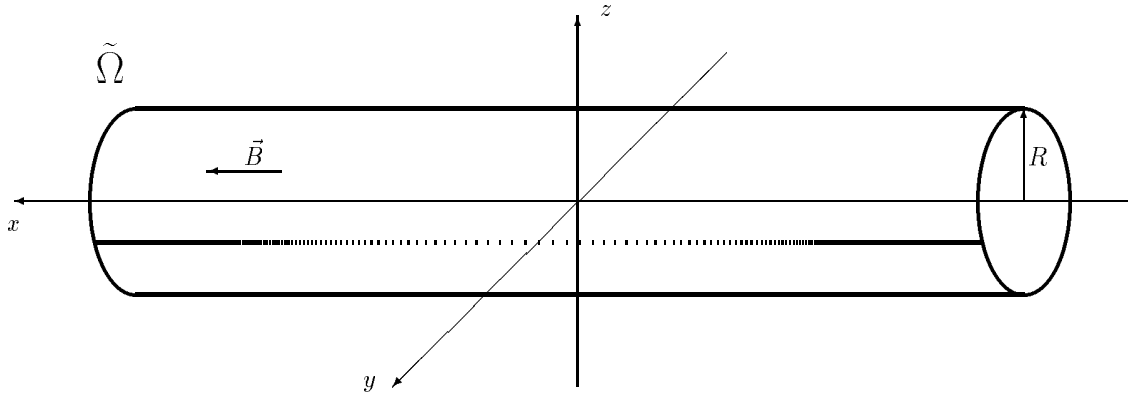}
\caption{Cylindrical strip with a $\delta$ barrier in axial
         magnetic field.}\label{Cschm}
\end{center}
\end{figure}

\subsection{General considerations}
The configuration space is sketched on Figure~\ref{Cschm} where we
indicate also how the coordinate system is chosen. In these
coordinates we have
\begin{equation}
  \tilde\Omega:=\{(x,y,z)\in\Real^3 |\ y^2+z^2=R^2\}\,.
\end{equation}
Choosing the gauge so that the electromagnetic potentials fulfill
$\varphi(\vec x)\equiv 0$ and $\vec A(\vec x)=\frac{1}{2}\vec
B\times\vec x$, the Hamiltonian acquires the form
\begin{equation}\label{CHamiltonian}
  \tilde H_{\alpha}:=(-i\nabla+\vec A)^2
  =-\Delta-i\frac{B}{2}(y\partial_z-z\partial_y)+\frac{B^2
  R^2}{4}
\end{equation}
away of the barrier, where $B:=|\vec B|$; as before we put
$\hbar=2m=1$, and also $e=-1$ having in mind an electron. The
subscript $\alpha$ indicates the real function which defines the
shape of the barrier as in the strip waveguide situation; it will
enter the boundary condition (\ref{newCbc}) below. The vector
potential has obviously the the angular component
$A_{\varphi}\equiv A$ and it equals
\begin{equation}\label{flux}
 A=\frac{1}{2}B R =\frac{\phi}{2\pi R},
\end{equation}
where $\phi$ is the magnetic flux through the cylinder. Recall
that in the rational units we use here, the natural unit of the
magnetic flux is $\phi_0:= (2\pi)^{-1}$. Since we deal with a
quantum system living on a surface it is natural to ``unfold" it
and to study~(\ref{CHamiltonian}) on a planar strip with
appropriate boundary conditions, namely
\begin{equation}\label{newCbc}
    \psi(u,0+)=\psi(u,2\pi R-)=:\psi(u,0), \quad
    \psi_v(u,0+)-\psi_v(u,2\pi R-)=\alpha(u)\psi(u,0),
\end{equation}
where the subscript denotes again a partial derivative. To this
aim, we introduce the unitary transformation
$U:\sii(\tilde\Omega)\to\sii(\Real\times[0,2\pi R],du dv)$ by
\begin{equation}
  (U\psi)(u,v):=\psi(u,R\cos\frac{v}{R},R\sin\frac{v}{R})\,,
\end{equation}
which maps $\tilde\Omega$ onto the strip
$\Omega:=\Real\times[0,2\pi R]$; the operator $\tilde H_\alpha$ is
then unitarily equivalent to
\begin{equation}\label{newCHamiltonian}
  H_{\alpha}:=U\tilde H_{\alpha} U^{-1}
  =-\partial_u^2+(-i\partial_v+A)^2\,,
\end{equation}
with the domain
\begin{equation}\label{CDomain}
  D(H_{\alpha}):=\left\{
  \psi\in\sobii(\Omega)\,|\ \forall u\in\Real:\; {\rm b.c.\;}
  (\ref{newCbc})\; {\rm are\; satisfied}\,
  \right\}.
\end{equation}
We will need also the quadratic form $q_{\alpha}$ associated with
$H_{\alpha}$. Its domain is $D(q_{\alpha}):=\sobi(\Omega)$ and
\begin{eqnarray}
  q_{\alpha}[\psi] &:=& \int_{\Omega}|\nabla\psi|^2(u,v) du dv
  +\int_{\Real}\alpha(u)|\psi(u,0)|^2 du\nonumber\\
  && - 2i A\int_{\Omega}(\bar\psi\partial_{v}\psi)(u,v) du dv
  +A^2\int_{\Omega}|\psi|^2(u,v) du dv\,.
\end{eqnarray}
As a comparison operator we employ again the one with
$\alpha(u)\!=\!\alpha\!=\!const$ when we can solve the
Schr\"odinger equation by separation of variables. We denote by
$\{\nu_n\}_{n=1}^{\infty}$ and $\{\chi_n\}_{n=1}^{\infty}$ the
(properly ordered) sequences of the transverse eigenvalues and the
corresponding eigenfunctions, respectively. Since our system is
now more complicated due to the presence of the magnetic field, we
have to distinguish several possibilities: \\ [2mm]
{\bf 1.}\quad \underline{No barrier}, $\alpha=0$:
\begin{equation}\label{CEigenfunctions1}
  \forall\ell\in\Int:\qquad\tilde\chi_{\ell}(v)=\frac{1}{\sqrt{2\pi R}}
  \ e^{i\frac{\ell}{R}v}\,.
\end{equation}
The tilde marks the eigenfunctions  corresponding to the
eigenvalues $(\frac{\ell}{R}+A)^2$; to get $\{\nu_n\}$ one has to
arrange thee latter into a ascending sequence. The respective
eigenfunctions will be then denoted as $\{\chi_n\}$.\\ [2mm]
{\bf 2.}\quad $\alpha\not=0$ \quad and \quad $2 R A\not\in\Nat$:
\begin{equation}\label{CEigenfunctions}
  \forall n\in\Nat\setminus\{0\}:\qquad
  \chi_n(v)=N_n\, e^{-i A v}\left(e^{i\sqrt{\nu_n}v}
  -\frac{e^{i 2\pi R A}-e^{i 2\pi R\sqrt{\nu_n}}}
        {e^{i 2\pi R A}-e^{-i 2\pi R\sqrt{\nu_n}}}\,
  e^{-i\sqrt{\nu_n}v}\right)\,,
\end{equation}
where $N_n$ denotes the normalization factor chosen to make
$\chi_n$ a unit vector in $\sii(0,2\pi R)$,
\begin{equation}
\begin{array}{c}
  |N_n|^2:=
  \frac{\sqrt{\nu_n}\ [1-\cos 2\pi R (\sqrt{\nu_n}+A)]}
  {4\pi R\sqrt{\nu_n}(1-\cos 2\pi R A\cos 2\pi R\sqrt{\nu_n})
  +2\sin 2\pi R\sqrt{\nu_n}(\cos 2\pi R A-\cos 2\pi
  R\sqrt{\nu_n})},
\end{array}
\end{equation}
and the increasing sequence $\{\nu_n\}$ arises from the spectral
condition
\begin{equation}\label{Cspc}
  -\alpha=2\ell\left(\cot 2\pi R\ell
  -\frac{\cos 2\pi R A}{\sin 2\pi R\ell}\right).
\end{equation}
In analogy with Lemma~\ref{approxbound}(a) we have
$\sqrt{\nu_n}\in\frac{1}{2 R}(n\!-\!1,n)$ for any
$n\in\Nat\setminus\{0\}$. \\[2mm]
{\bf 3.}\quad $\alpha\not=0$ \quad and \underline{integer flux},
$2 R A\in 2\Nat$: \\
The transverse eigenfunctions are of the
form~(\ref{CEigenfunctions}), while the spectral
condition~(\ref{Cspc}) changes to
\begin{equation}\label{Cspceven}
  \alpha=2\ell\tan\pi R\ell\,.
\end{equation}
Moreover, for $\ell\in\frac{1}{R}\Nat$ we always get the trivial
solutions
\begin{equation}\label{CEigenfunctionstriv}
  \chi_n^{triv}(v)=\frac{1}{\sqrt{\pi R}}\,e^{-i A v}
  \sin \ell v\,,
\end{equation}
which are independent of $\alpha$. The roots $\{\tilde\nu_n\}$
of~(\ref{Cspceven}) satisfy the estimates
$\sqrt{\tilde\nu_1}\in\frac{1}{2 R}(0,1)$ and
$\sqrt{\tilde\nu_n}\in\frac{1}{2 R}(2 n\!-\!3,2 n\!-\!1)$ for
$n\in\Nat\setminus\{0,1\}$. \\ [2mm]
{\bf 4.}\quad $\alpha\not=0$ \quad and \underline{half-integer
flux}, $2 R A\in 2\Nat+1$: \\
As in the two preceding cases the transverse eigenfunctions are
still~(\ref{CEigenfunctions}) but the spectral
condition~(\ref{Cspc}) changes now to
\begin{equation}\label{Cspcodd}
  -\alpha=2 \ell\cot\pi R \ell\,.
\end{equation}
The wave functions of the trivial solutions, $\ell\in\frac{1}{2
R}(2\Nat\!+\! 1)$ are~(\ref{CEigenfunctionstriv}) again and the
nontrivial roots of~(\ref{Cspcodd}) satisfy
$\sqrt{\tilde\nu_n}\in\frac{1}{2 R}(2 n\!-\!2,2 n)$ for all
$n\in\Nat\setminus\{0\}$. \\
\begin{rem}\label{Crems1}{\rm
The integer and half-integer values here refer to the natural flux
unit mentioned above. The essential instrument for proving the
existence of bound states is the requirement $\chi_1(0)\not=0$
(compare, \eg, to Eq.~(\ref{Q=}) above). In the absence of a
barrier, $\alpha=0$, the eigenfunctions are always positive
(see~(\ref{CEigenfunctions1})) and $\chi_1(0)\not=0$ holds for
$\alpha\leq\alpha_m:=-\frac{2\sin^2\pi R A}{\pi R}$. In the case
of a (half-)integer flux we have to exclude the trivial
solutions~(\ref{CEigenfunctionstriv}). It is an analogy of the
trivial-part exclusion described in~Remark~\ref{rem1}; the
difference is that the triviality now does not come from the
waveguide geometry, but rather from the magnetic field, \ie, an
external parameter. It is also clear that the ground-state
eigenfunction of the class (\ref{CEigenfunctions}) can vanish at
the barrier only if $\alpha>0$ and the flux is half-integer.}
\end{rem}

    In analogy with Lemma~\ref{approxbound} we have
\begin{lemma}\label{Capproxbound}
    Suppose that $2RA\not\in 2\Nat+1$ or $\alpha\leq0$.
Then the function $\alpha\mapsto\nu_1(\alpha)$ is strictly
increasing and continuous.
\end{lemma}
    The ``unperturbed" Green's function is the same as in the
case of double waveguide (\cf~(\ref{green0})); one has only to
substitute the present transverse eigenfunctions and eigenvalues.

After this preliminary we can easily derive sufficient conditions
under which a local perturbation of the barrier coupling parameter
induces existence of bound states. The argument mimics that of
Theorem~\ref{existence}, the only difference being an additional
requirement of the magnetic field.
\begin{thm}\label{Cexistence}
    Assume {\em (i)}  $\;\alpha\!-\!\alpha_0\in\si_{loc}(\Real)$,
    \\[1mm]
 {\em (ii)}$\;\alpha(u)\!-\!\alpha_0={\cal O}(|u|^{-1-\varepsilon})$
for some $\varepsilon>0$ as $|u|\to\infty$,\\[1mm]
 {\em (iii)} the flux is not half-integer, $2RA\not\in 2\Nat+1$,
if $\alpha_0>0$. \\[1mm]
 Then $H_{\alpha}$ has at least one
isolated eigenvalue below its essential spectrum,
$\sigma_{ess}(H_{\alpha})= [\nu_1(\alpha_0),\infty)$, provided
$\int_{\Real}(\alpha(u)\!-\!\alpha_0)du<0$.
\end{thm}

\begin{figure}[!htb]
\begin{center}
\includegraphics[width=\textwidth]{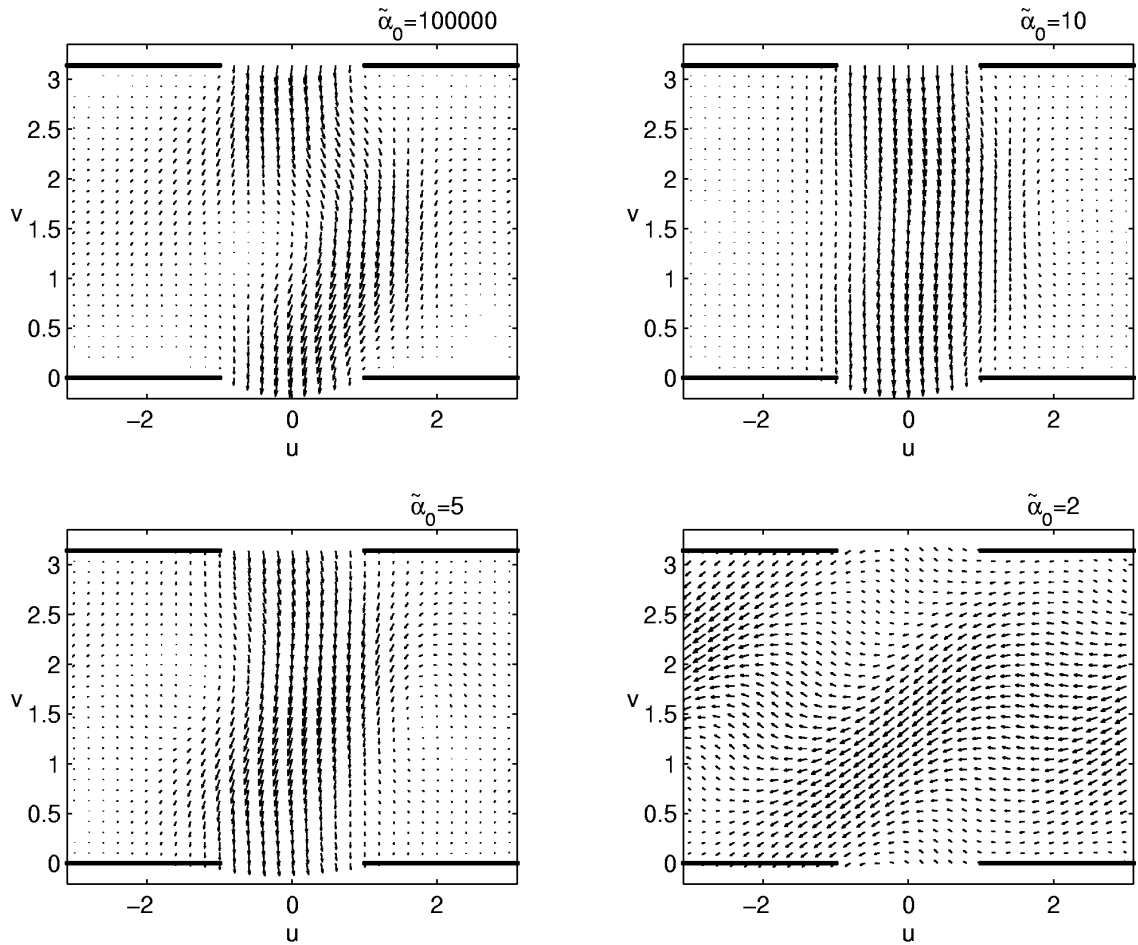}
\end{center}
\caption{The quantum probability flow on the cylinder surface for $\tilde A=0.5$, 
         $\tilde k=1.705$ and $\tilde\alpha_1=10^{-5}$.}\label{CfigFlow}
\end{figure}

\subsection{An example}
We shall illustrate the above consideration on a
``rectangular-well" example analogous to that of
Section~\ref{Sec.Example}. Most of the argument proceeds as there,
one has just to use the different eigenfunctions and to recompute
for them the overlap integrals:
\begin{eqnarray}\label{Coverlap1}
  (\chi_m,\phi_n) &=& \frac{8\,\bar N_m^0 N_n^1}
  {(e^{-i 2\pi R A}-e^{i 2\pi R\sqrt{\nu_m^0}}\ )
  (e^{i 2\pi R A}-e^{-i 2\pi R\sqrt{\nu_n^1}}\ )\,
  (\nu_m^0-\nu_n^1)}\cdot\nonumber\\
  &&\cdot\Bigl[\sqrt{\nu_m^0}\sin2\pi R\sqrt{\nu_n^1}
  \ (\cos 2\pi R A-\cos 2\pi R\sqrt{\nu_m^0}\ )\nonumber\\
  &&\ -\sqrt{\nu_n^1}\sin 2\pi R\sqrt{\nu_m^0}
  \ (\cos 2\pi R A-\cos 2\pi R\sqrt{\nu_n^1}\ )\Bigr]
\end{eqnarray}
for $\alpha_0\not=0\not=\alpha_1$ with $m,n\in\Nat\setminus\{0\}$,
and
\begin{equation}\label{Coverlap2}
  (\chi_m,\phi_{\ell})=\frac{4\,\bar N_m^0\,\sqrt{\nu_m^0}}
  {\sqrt{2\pi R}\,\left(\nu_m^0-(\frac{\ell}{R}+A)^2\right)}\
  \left(\sin 2\pi R A+i\cos 2\pi R\sqrt{\nu_m^0}\,\right)
\end{equation}
for $\alpha_0\not=0$, $\alpha_1=0$ with $m\in\Nat\setminus\{0\}$,
$\ell\in\Int$. Note that in the latter case one has to substitute
the {\em ordered} basis $\{\phi_n\}_{n=1}^{\infty}$ (together with
the corresponding eigenvalues) into Ansatz~(\ref{Ansatz}) to make
the numerical procedure of cut-off approximations convergent.

    Unless $\alpha_1<0$ we have to exclude
here the possibility $2 R A\in 2\Nat\!+\!1$ again (\cf,
\eg,~(\ref{estg})). Next we can restrict our attention only to the
situation when $2 R A\in [0,1)\cup(1,2)$ because $A$ appears in
the overlap integrals~(\ref{Coverlap1}), (\ref{Coverlap2}) and in
the spectral condition~(\ref{Cspc}) as an argument of the periodic
functions $\sin$, $\cos$; the integrals and the transverse
eigenvalues are the only quantities which affect the
equations~(\ref{C.a=0}) and~(\ref{C.a=f}).

    In fact, we can take $2 R A$ from $[0,1)$ only because
the replacement $2 R A\mapsto 2-2 R A$ in~(\ref{Coverlap1})
and~(\ref{Coverlap2}) (the spectral condition~(\ref{Cspc}) does
not change at all) is equivalent to the exchange $A\mapsto -A$
which coincides with the conjugation of
Hamiltonian~(\ref{newCHamiltonian}). It is well known fact that
such an operator has the same energies while the corresponding
eigenfunctions are given by a simple conjugation.

    As an example, the evolution of the quantum probability flow
w.r.t. $\alpha_0$ is illustrated on Figure~\ref{CfigFlow}.


\bigskip\bigskip
\listoffigures

\end{document}

%% file: fig01.tex
\newbox\stroke\setbox\stroke\hbox{}        
\newdimen\spac\spac=.5mm
\newdimen\xm\xm=90mm
\newdimen\xcntr\xcntr=45mm
\newcount\multipler\multipler=1
\def\dott{\vrule height .5mm depth 0mm}
\def\fill{
\setbox\stroke=\hbox{\unhbox\stroke\hskip\spac\dott}
\multiply\spac by \multipler\divide\spac by 1000
\ifdim\wd\stroke>\xcntr\multipler=950\fi
\ifdim\wd\stroke<\xcntr\multipler=1050\fi
\ifdim\wd\stroke<\xm\expandafter\fill\fi}
\fill
\def\barrier{\copy\stroke}
\setlength{\unitlength}{1mm}
\begin{picture}(120,30)(-1,10)
  \put(10,14){\rule{\xm}{1mm}}
  \put(10,30){\rule{\xm}{1mm}}
  \put(10,19.8){\barrier}
  \put(0,20){\line(1,0){12}}
  \put(100,20){\vector(1,0){10}}
  \put(55,8){\vector(0,1){30}}
  \put(110,17){$x$}
  \put(57,38){$y$}
  \put(48.5,10.1){$-d_{2}$}
  \put(51.5,32.5){$d_{1}$}
\end{picture}